\documentclass[aps,reprint,pre]{revtex4-1}
\usepackage[usenames]{xcolor}
\usepackage{amsfonts, amsmath, amssymb}
\usepackage[pdftex]{graphicx}
\DeclareMathOperator{\sign}{sgn}
\DeclareMathOperator{\arccot}{arccot}
\newcommand{\TSP}{{\mathrm{TSP}}}

\begin{document}
\title{Complexity in genetic networks: topology vs. strength of interactions}
\author{Mikhail Tikhonov}
\author{William Bialek}
\affiliation{Joseph Henry Laboratories of Physics, and Lewis--Sigler Institute for Integrative Genomics, Princeton University, Princeton, NJ 08540, USA}
\begin{abstract}
Genetic regulatory networks are defined by their topology and by a multitude of continuously adjustable parameters.  Here we present a class of simple models within which the relative importance of topology vs. interaction strengths becomes a well--posed problem.  We find that complexity---the ability of the network to adopt multiple stable states---is dominated by the  adjustable parameters.  We comment on the implications for real networks and their evolution.
\end{abstract}

\maketitle

What we think of as the state of a living cell is determined largely by the concentrations of various proteins.  But the instructions for making the proteins are encoded in DNA, and the rates at which this information is read out from different genes is determined, in part, by the concentration of other proteins.  Thus, there is a network of interactions in which genes encode proteins, and proteins control the reading out of the genes.  Such ``genetic  regulatory networks''  are not the whole story of how cells control their states, but this is a good starting point, and the image of cellular states as being the states of an interacting network certainly has shaped quite a bit of thinking about cellular function \cite{GRNs}.

During embryonic development of multicellular organisms, the states of the relevant genetic networks are thought to encode the body plan of the adult \cite{lawrence_92,gerhardt+kirschner_97}, and so the ability of the network to adopt a richer set of states corresponds to building a more complex organism.  What is it about the network that controls this complexity?  How do the changes in DNA sequence allow the emergence of greater complexity over the course of evolutionary history \cite{carroll_05}?

Much of what we know about the structure of genetic networks comes from classical genetics---we see what happens when a mutation knocks out one element of the network.   Such experiments lead to information about the topology of the network:  the protein encoded by gene A represses the read out of genes C, and, activates the read out of gene D, and does nothing to genes B and F.  It is much more difficult to measure the strength of these interactions. Perhaps because of this experimental situation, there has been a considerable focus on network topology itself as a determinant of biological function.

Ideas about network topology include several themes.  One approach aims at a statistical characterization of network topologies, focusing on the distribution of the number of connections to a single node (degree distribution) \cite{barabasi} or the  presence of local motifs in which patterns of connections among small numbers of genes are over--represented \cite{lee+al_02,milo+al_02}.  Another idea is that relatively small changes in DNA sequence in the regions where proteins bind and regulate the expression of genes can change the effective topology of the network, and thus there is a path for topology to evolve quickly, without changing the number or identity of genes \cite{carroll_05,early}.   Finally there is the idea that the difficulty of defining interactions strengths is a problem not only for us but for the cell itself, and hence that important cellular functions must be ``robust'' against variations in these parameters, or equivalently against changes in the absolute numbers of all the relevant protein molecules \cite{robust,dassow_00,bialek_12}; taken literally, this means that function must be  encoded in topology alone.   

While the focus on network topology is widespread, there are alternatives.  Rather than being irrelevant, parameters could be optimized to transmit the maximum amount of information through a network \cite{infomax}, to achieve the maximum signal--to--noise ratio for weak signals \cite{snrmax}, or to insure that events occur in a precisely timed sequence \cite{ronen+al_02}.  The experimental situation is challenging, but it ought to be possible to define a class of models within which the relative contributions of topology and parameters can be dissected completely.  

To make the comparison of topology and interaction strengths a well posed problem, we have to address two issues.  First, we have to say what we mean by complexity.  Second, we have to define  a measure on the space of interaction strengths.  The claim, for example, that ``typical'' parameter values lead to certain behaviors depends on the shape of the distribution over parameter space.  Since interactions are determined by the rates or equilibrium constants for biochemical reactions (e.g., binding of a protein to a site along the DNA \cite{bintu}), these parameters are continuous and  exponentially sensitive to perhaps more natural parameters such as binding energies; the question of what constitutes a natural measure on such a space is not trivial.   

Here we introduce a highly simplified model in which these issues have a natural formulation, and in particular where the continuous parameter space breaks into discrete sectors with equal weight, so that varying parameters and varying topology both become matters of enumeration.   Within this model we will see that complexity is dominated by the choice of parameters, and that it is very difficult to evolve greater complexity by changing topology without optimizing parameters.

\begin{figure}[t]
\includegraphics[width = 0.9 \linewidth]{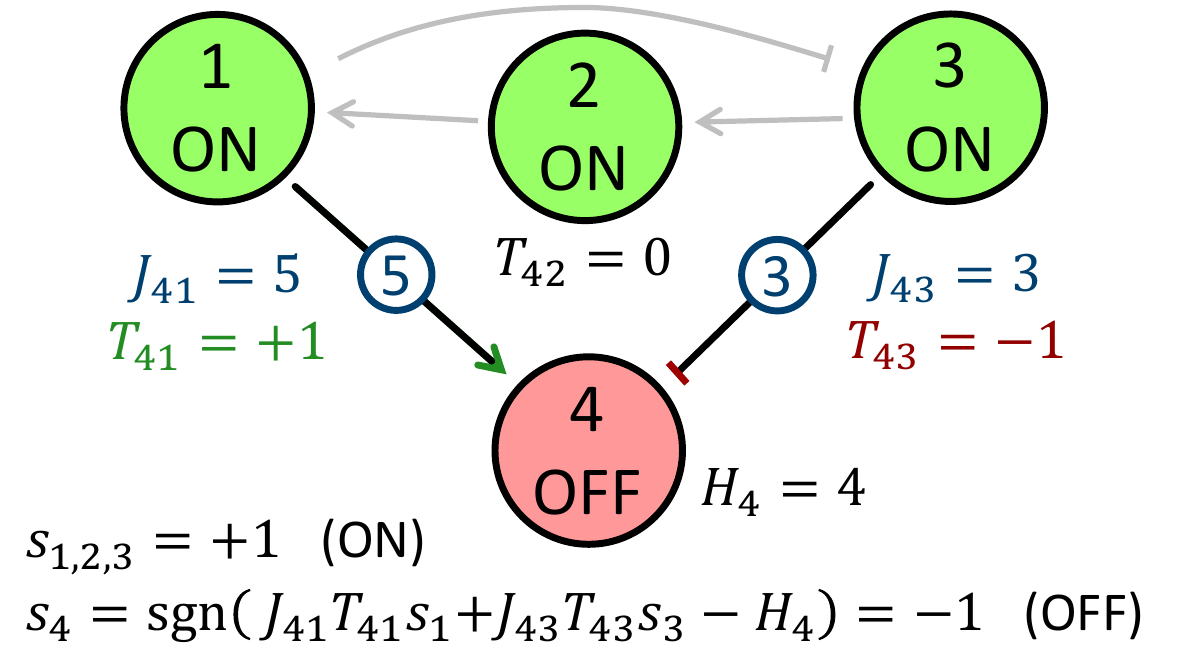}
\caption{In this example, we focus on node 4 to illustrate our notations and the activation rule. Node 4 is activated by node 1 with strength 5 and repressed by node 3 (strength 3). It is constitutively repressed with strength $H_4$=4. The regulatory rule dictates node 4 to remain ``off.''    \label{fig:notation}}
\end{figure}

To make a simplified model, we imagine that every gene $\rm i$ has a binary state, $s_{\rm i} =\pm 1$, where $s_{\rm i} = +1$ indicates that protein encoded by gene $\rm i$ is being synthesized, and thus is present at a relatively high concentration, while $s_{\rm i} = -1$ indicates that this protein is at near zero concentration.  The state is determined by inputs from other proteins, and we assume that these inputs add; the gene is  ``on'' if the total input exceeds a threshold:
\begin{equation}
s_{\rm i} = \sign\left[ \sum_{\rm j} {\hat J}_{\rm ij} s_{\rm j} - H_{\rm i}\right] .
\label{act1}
\end{equation}
The matrix  $\hat J$ encodes both the topology of the network and strength of the interactions.  To separate these we write
${\hat J}_{\rm ij} = J_{\rm ij} T_{\rm ij}$,
where the elements of the topology matrix $T_{\rm ij}$ are assigned values $+1$, $-1$, or $0$ depending on whether the protein encoded by gene $\rm j$ activates, represses, or does nothing to gene $\rm i$.  The interaction matrix $J_{\rm ij}$ then can be assigned all positive elements.  In a similar spirit, we write $H_{\rm i} = c_{\rm i} h_{\rm i}$, where $c_{\rm i} = \pm 1$ and $h_{\rm i} \geq 0$; $c_{\rm i} = +1$ means that gene $\rm i$ would be ``on'' in the absence of inputs (``constitutively active,'' in biological terms), and conversely for $c_{\rm i} = -1$.  Thus, Eq (\ref{act1}) becomes
\begin{equation}
s_{\rm i} = \sign\left[ \sum_{\rm j} {T}_{\rm ij} J_{\rm ij} s_{\rm j} - c_{\rm i} h_{\rm i}\right] .
\label{act2}
\end{equation}
The network is defined by its  topology $\Gamma \equiv \{T_{\rm ij}, c_{\rm i}\}$ and its continuous parameters or weightings  $W\equiv \{J_{\rm ij} , h_{\rm i}\}$.

A configuration of on/off states $\{s_{\rm i}=\pm 1\}$ that satisfies Eq (\ref{act2}) at every node will be called a \emph{solution} of the network; solutions  describe configurations of static gene activity patterns within a single cell. Their number may correspond, for example, to the number of cell types this network can encode during development, and so is a natural measure of network complexity. In the language of neural networks \cite{hopfield_82,amit_89}, we can call it the \emph{capacity} of the network, $c(\Gamma, W)$. It can also be thought of as a measure of information processing capability: for example, a network possessing just two solutions $\{s_{\rm i}\}$ and $\{-s_{\rm i}\}$ can be seen as  taking one input (state of node $s_1$) and setting the state of other nodes to well--defined values that depend on this input, while a network with more solutions is capable of distinguishing more combinations of inputs and adjusting the outputs accordingly, and so performs a more complex computation.   Our task, then, is to compute $c(\Gamma, W)$  \cite{boolean,kauffman_69}.

Although the parameters $\{J_{\rm ij}, h_{\rm i}\}$ are continuous and, in principle, unbounded, if we are only interested in the {\em solutions} of the network, there is a natural compact geometry to the parameter space, and this geometry also breaks into discrete subspaces.  To see this, let's denote by $\vec w_{\rm i}$ the vector of all parameters that ``feed'' into gene $\rm i$, $\vec w_{\rm i} \equiv \{J_{{\rm i}1}, J_{{\rm i}2}, \cdots , h_{\rm i}\}$.  With $N$ genes, there are $N$ separate vectors $\vec w_{\rm i}$, and together these vectors define the parameter space of the model.    But Eq (\ref{act2}) has a symmetry, where the states of the system are invariant under independent scaling of the parameters feeding into each node, $\vec w_{\rm i} \rightarrow \alpha_{\rm i}\vec w_{\rm i}$ \cite{noise}.    Thus we can choose $|\vec w_{\rm i}| =1$ for each gene $\rm i$, so that the relevant parameter space is a direct product of positive segments of unit spheres.

Once we realize that the relevant parameter space is the portion of the unit sphere, there is a natural measure, namely the uniform distribution.  But we can do more, because whole sectors of the parameter space produce the same solutions.  We know this has to be true because Eq (\ref{act2}) says that each node computes a Boolean function of its inputs, and there is only a finite number of Boolean functions with $N$ inputs.  Further, our model generates only a very small subset of these, the perceptrons \cite{perceptron}.

\begin{figure}[b]
\includegraphics[width = 0.47 \textwidth]{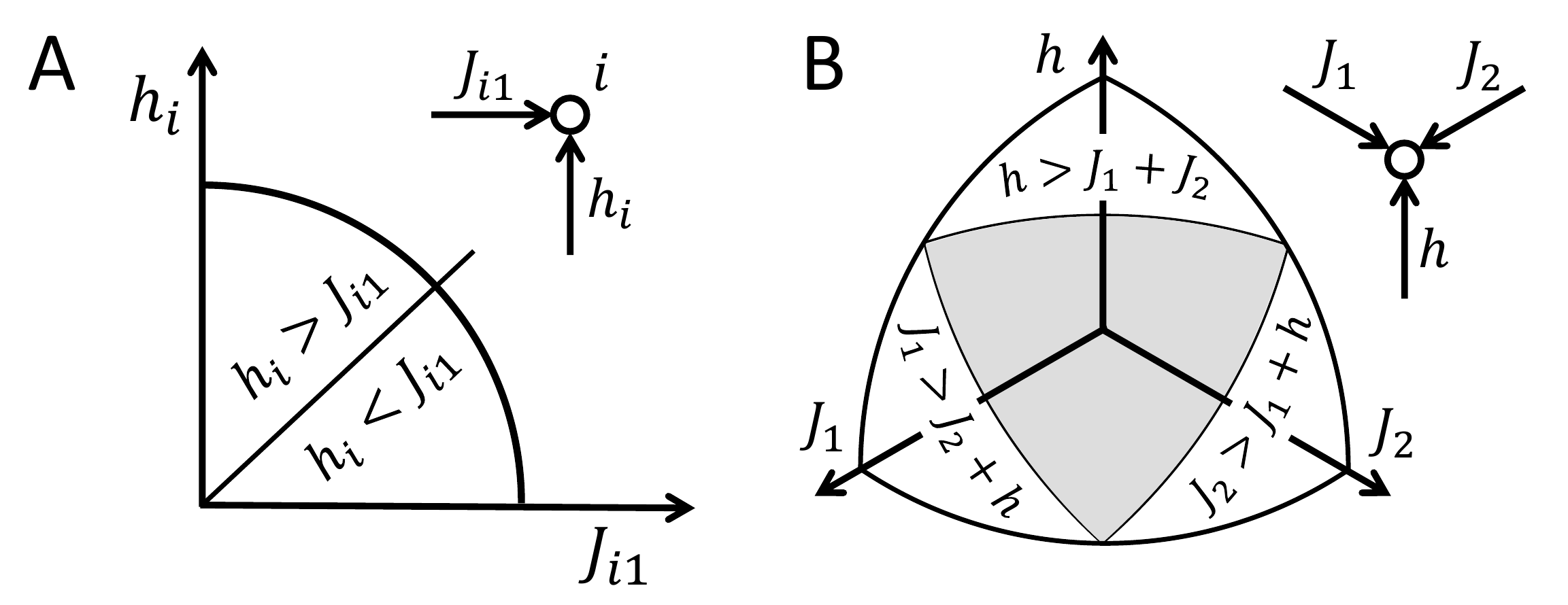}
\caption{Parameter space splits into discrete weighting sectors. \textbf{A:} The relative strength of two inputs to node $\rm i$, $J_{{\rm i}1}$ and $h_1$,  is parametrized by a point on a circle. There are two sectors: $h_1>J_{{\rm i}1}$ and $J_{{\rm i}1}>h_1$. \textbf{B:} For three inputs, the parameter space is a 2-sphere and splits into 4 sectors: three in which one input dominates, and the unique non--dominating sector (shaded).   \label{fig:weightSectors}}
\end{figure}

In the simplest case, shown in Fig \ref{fig:weightSectors}a, a gene $\rm i$ receives input from one other gene (with strength $J_{{\rm i}1}$) and compares this to a local threshold $h_{\rm i}$; now the ``unit sphere'' is just a quarter circle.  But if $J_{{\rm i}1}<h_{\rm i}$, then gene $\rm i$ will have the same state, $s_{\rm i} = \sign c_{\rm i}$, no matter what the state of the input $s_{\rm 1}$ might be.  On the other hand, if $J_{{\rm i}1}>h_{\rm i}$, then the state of gene $\rm i$ is determined uniquely by the state of the input, $s_{\rm i} = \sign(T_{{\rm i}1}s_1)$.   These sectors have the same weight under the uniform distribution.

The simplest two--dimensional case also alerts us to a problem, namely that some combinations of parameters aren't very interesting.  In this case, there are two possibilities, and in one of them the gene $\rm i$ is essentially uncoupled from the network.  In the other case, gene $\rm i$ is completely redundant with gene $1$.  Somehow neither of these cases sounds much like a ``network.''

The next simplest case is where gene $\rm i$ takes two inputs and compares their sum to a threshold (``in degree''~2).  Now the vector $\vec w_{\rm i}$ is three dimensional, so that the relevant space is a segment of the familiar unit two--sphere embedded in three dimensions, shown in Fig \ref{fig:weightSectors}b.  Again the continuous space breaks into discrete regions,  and within each region the solutions of the network are the same.  There are four regions, and three of them are uninteresting in the same way that we saw for the two dimensional case.  In one sector, the state of gene $\rm i$ is dominated by the threshold $h_{\rm i}$, and so this node is effectively not attached to the network, since its state is independent of the state of all other nodes.  In two other sectors, one of the interactions $J_{\rm ij}$ is so large that it dominates all other inputs, and hence gene $\rm i$ is completely redundant with one other gene $\rm j$. It is only in the remaining fourth sector where the state of gene $\rm i$ depends in a nontrivial way on the combination of its inputs.

The picture of dominating sectors in Fig \ref{fig:weightSectors}b---regions of parameter space where a single input dominates, so that one node becomes either decoupled from the network or completely redundant with one other node---is the quantitative expression of the intuition that making interactions continuously weaker is not obviously separable from changing topology by erasing these interactions altogether.  In the dominating sectors, interactions can be erased without changing the set of solutions in the network.  In this sense, parameter settings in such dominating sectors are functionally equivalent to networks with simpler topology (fewer connections), and so if we want to compare the role of continuous parameters to that of network topology, we should exclude these regions of parameter space (see also Appendix \ref{app:IWCsectors}).

With three inputs to each node,  the vector $\vec w_{\rm i}$ is four dimensional, and there are twelve discrete sectors of the sphere that generate different solutions.  Four of these have a single dominating weight, and thus will be excluded.  The remaining eight sectors have equal probability under the uniform distribution on the sphere (Appendix \ref{app:IWCvolumes}).  Thus, if we consider networks in which each gene is influenced by three inputs, there are $8^N$ distinct networks for a given topology, and we can do exhaustive enumerations up to values of $N$ that are typical of real genetic networks.  Already with four inputs there are $(76)^N$ networks for each topology, so we will confine our attention to the case of three inputs.

\begin{figure}[b]
\includegraphics[width = 0.9 \linewidth]{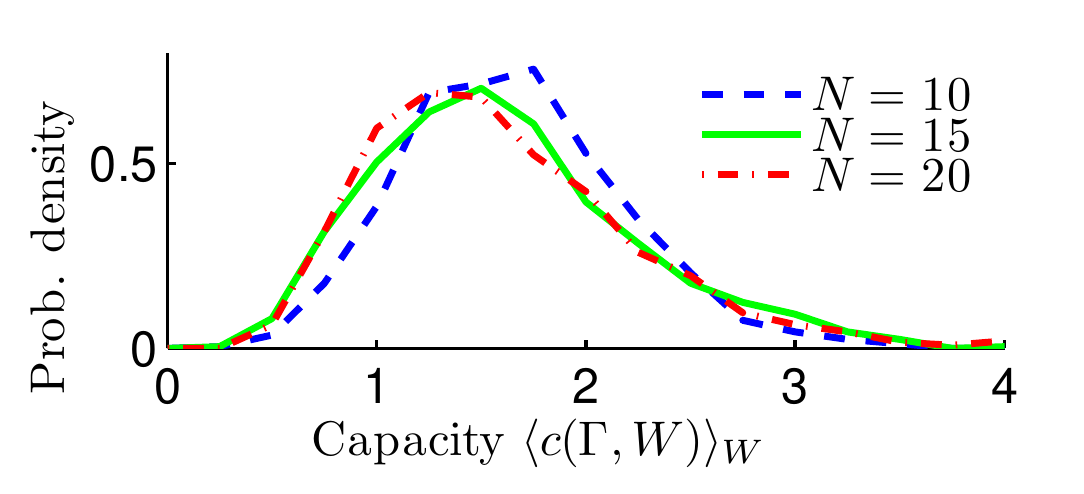}
\caption{Distribution of average complexity over non--dominating weightings, calculated for 1000 random topologies. The average complexity does not grow with $N$. \label{fig:changeN}}
\end{figure}

Let us start with $N$ genes connected in a particular topology, and then choose parameters at random from the $8^N$ sectors described above.  For each network we can measure the capacity, or number of distinct solutions to Eq (\ref{act2}), and then average over parameters at fixed topology.  In Fig \ref{fig:changeN} we show the distribution of this mean complexity across topologies in which all $N$ genes are repressors. Surprisingly, the average complexity is independent of $N$. This may appear counterintuitive: we expect larger graphs to be capable of storing more patterns, but they also have more weightings with few or no solutions. This result can be demonstrated analytically with a mean-field argument that holds independently of our simplifying assumptions such as constant in--degree (see Appendix \ref{app:Computation}). It shows that, quite generally, a larger network is not automatically more complex; rather, it has the potential for high complexity, but only realizes this potential with a careful choice of weights.

\begin{figure*}[!htbp]
\includegraphics[width = \textwidth]{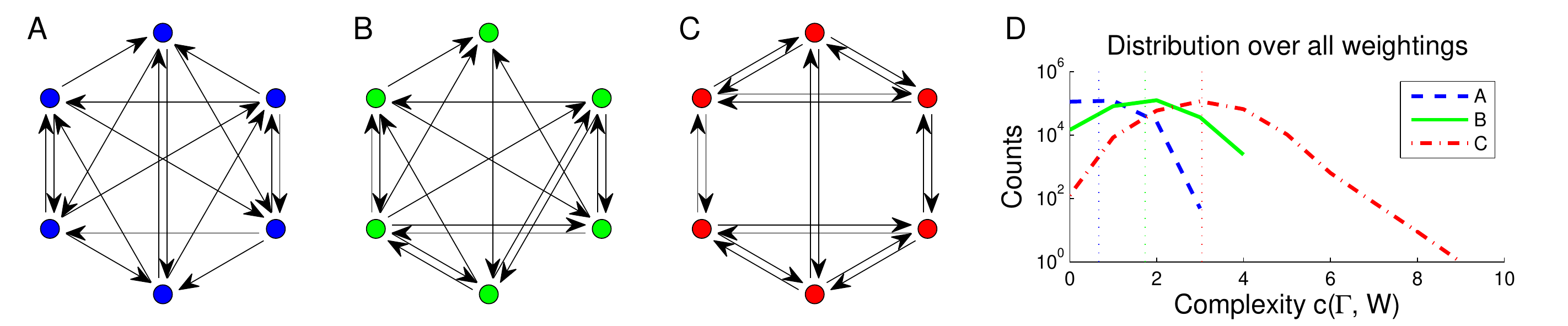}
\caption{\textbf{A-C:} $N=6$ graphs with lowest, median and highest average complexity. \textbf{D:} Distributions of complexity of these graphs over the choice of weighting overlap significantly. Vertical lines show the average complexity (0.7, 1.7 and 3.0, respectively). Even the ``worst'' graph (A) can be optimized to attain the typical complexity of the best topology (C). \label{fig:N6}}
\end{figure*}

To highlight the difference between average and attainable complexity, we calculate, for a given topology, the distribution of $c(\Gamma, W)$ over all weightings $W$. For $N=6$ we can enumerate all topologies; Figure~\ref{fig:N6} shows three examples with the lowest, typical and highest average complexity, and the corresponding distributions. We observe that the distributions overlap considerably, and the typical realizations of even the best topology are routinely outperformed by ``lesser'' topologies when their weights are optimized. This persists for larger $N$: the best $N=9$ topology found by a targeted search has average complexity $\max_\Gamma \langle c(\Gamma ,W)\rangle_W = 5.26$. This is far in the tail of the distribution: uniform sampling of 1000 topologies gives an average complexity of only $\langle c(\Gamma, W)\rangle_{\Gamma,W} = 1.7 \pm 0.5$. However, optimizing weights of random topologies gives higher complexity in 85\% of the samples. In other words, if one were forced to pick only one feature to optimize, either weights or topology but not both, optimizing weights is the better strategy.

Before interpreting these results, we should be careful about our definition of complexity,  since highly disconnected graphs can achieve high capacity without being complex in any intuitive sense of the word. For example, the maximal capacity of a network with $M$ mutually repressing pairs (Fig ~\ref{fig:diagram}A) grows exponentially,  $\max_W c(\Gamma, W)=2^M$. We note, however, that the maximal capacities that we find for $N$--gene networks ($\geq27$ for $N=9$, $\geq 54$ for $N=10$) are larger than $2^{\lfloor N/2\rfloor}$, i.e. storing patters in a distributed way is more efficient than splitting the network into many disconnected components. Therefore, the highest capacity networks are not trivially so. Further, in the set of all graphs, disconnected topologies are extremely rare and can be expected to provide but a small correction to the averages. One can construct a better definition, quantifying complexity of a network as the diversity of causal relations in the set of its solutions (Appendix \ref{app:TSP}). This definition naturally handles pathological cases and solution degeneracies, yet for connected graphs it is in excellent agreement with $c(\Gamma, W)$, which is much easier to compute.

Within the class of models that we have studied here, it is possible to dissect the contributions of topology and interaction strengths to network complexity. Unambiguously, interactions strengths are dominant:  starting from a random network, adjusting the strengths of regulatory interactions is  a better strategy for increasing its complexity than changing the topology or even adding new nodes \cite{note2}.  As in the brain, it seems that topology provides the potential for complexity, but parameters must be adjusted carefully to  realize this potential.

Maximally complex network function is accessible only in a small fraction of parameter space, but optimization does not lead to unique parameter settings.  Instead, there is a whole sector of the underlying continuous parameter space that produces the same results.  Thus, in these models, maximizing complexity leads to an interesting combination of tuning and tolerance.   Topologies with the greatest potential for complexity do have local structures in common with those identified in real networks \cite{lee+al_02,milo+al_02}, but it is an open question whether biasing the sampling of topologies toward these structures would, on average, enhance the complexity of network function.

Adjusting parameters is a more efficient method for increasing complexity, as illustrated in Fig \ref{fig:N6}, but it still is possible that the evolution of complexity is associated with changes in network topology \cite{carroll_05}.  If continuous parameters can evolve more rapidly than topologies can change, which seems plausible, then today's organisms may be dominated by networks that are near optimal given their topology.  In this scenario, today's more complex organisms must have networks with different topology than their less complex counterparts, but not because parameters are irrelevant---rather, topology becomes determining only once parameters have been optimized.

We  thank CP Broedersz, BB Machta, DJ Schwab, and NS Wingreen  for helpful discussions.  This work was supported in part by NSF grants CCF--0939370, PHY--0957573, and PHY--1022140, and by the WM Keck Foundation.

\appendix

\section{Weighting sectors}\label{app:IWCsectors}

Our model is defined by Eq \ref{act2},  
\begin{equation}\label{app:regulatoryRule}
s_{\rm i}=\sign \left[\sum_{\rm j} T_{\rm ij} J_{\rm ij}  s_{\rm j}-c_{\rm i} h_{\rm i}\right] ,
\end{equation}
with notation as explained in the main text. For technical reasons we will set  $T_{\rm ii}=0$, forbidding explicit auto--regulation \cite{auto}.
 We will continue to group the parameters of the network into the topology $\Gamma\equiv\{T_{\rm ij},c_{\rm i}\}$  and the weights $W\equiv\{J_{\rm ij}, h_{\rm i}\}$.

To formally define the weighting sectors, we first notice that the $-H_{\rm i}=-c_{\rm i}h_{\rm i}$ term in Eq \eqref{app:regulatoryRule} can be seen as an additional input from a constitutively expressed repressor (if $c_{\rm i}=+1$) or activator (if $c_{\rm i}=-1$), so we can think of input weights and the activation threshold in a unified way. Mathematically, we then have the following structure. Consider a network with a specified weighting, and a given configuration of node states $s_{\rm i}$ ($\pm 1 = \mathrm{on/off}$). Each link ${\rm i\rightarrow j}$ in the network is either satisfied or frustrated; the ``frustration state'' of a link will be denoted as $f_{\rm i\rightarrow j} \equiv s_{\rm i} s_{\rm j} T_{\rm ji}$ and takes a binary value, $f_{\rm i\rightarrow j}=\pm1$. Interpreting activation thresholds as constitutive activation or repression by a ``virtual'' node, the corresponding ``virtual'' link can also be satisfied or frustrated (the virtual link is satisfied for a constitutively activated node ($h<0$) that is \emph{on} or a constitutively repressed node ($h>0$) that is \emph{off}). For the rest of the appendices,
we will treat internal thresholds as weights of these additional ``virtual'' interactions, considering $h_{\rm i}$ as part of an extended matrix $\tilde J_{\rm ij}$; we will use the tilde as a reminder that the internal thresholds are treated as weights of additional links.

A ``weighting sector'' is a map that determines whether a particular combination of satisfied/frustrated input links is consistent with Eq \eqref{app:regulatoryRule}. The full parameter space is a direct product of the parameter spaces describing individual nodes, so to simplify notation, let us focus on one node $\rm i_0$. Denote $U(\rm i_0)$ the complete set of its inputs (genes that regulate $\rm i_0$, as well as the internal threshold), and $K+1$ their number:
\begin{equation}
U(\rm i_0)=\{{\rm j} \mid \tilde T_{\rm ij}\neq 0\} ,
\end{equation}
where $\tilde T$ expands $T$ to include the connections to virtual nodes that model the threshold.   Now, let $\vec w$ and $\vec t$ be the $(K+1)$-element vectors of strengths and signs of the interactions regulating gene $\rm i_0$ (the incoming links):
\begin{align}
\vec w &= \left. \tilde J_{\rm i_0j}\right|_{{\rm j}\in U(\rm i_0)} \in \mathbb{R}_+^{K+1}\\
\vec t &= \left. \tilde T_{\rm i_0j}\right|_{{\rm j}\in U(\rm i_0)} \in \pm1
\end{align}
For a fixed topology $\vec t$, the incoming weights $\vec w$ at node $\rm i_0$ define a Boolean function of $K+1$ arguments $\phi_{\vec w}\colon\{1, -1\}^{K+1}\mapsto \{1, -1\}$:
\begin{equation}
\phi_{\vec w}(b_1, b_2, \dots, b_{K+1})=\sign\left(\sum_{\rm j} b_{\rm j} t_{\rm j} w_{\rm j}\right).
\end{equation}
This function has the following interpretation: when applied to the set $\{f_{\rm j\rightarrow i_0}\mid{\rm j}\in U(\rm i_0)\}$, it tells us if this combination of satisfied/frustrated links is allowed by the regulatory rule at node $\rm i_0$. Note that this is \emph{not} the function that maps the states of input nodes into the state of the target node (the input/output function); using $\phi$ allows us to exhibit the symmetry between all $K+1$ inputs, whereas the input-output function must treat the internal threshold separately.

A ``weighting sector'' at node $\rm i_0$ is the equivalence class of vectors $\vec w$ that define the same Boolean function $\phi_{\vec w}$. The set of possible Boolean functions, which is much smaller than the set of all possible Boolean functions \cite{boolean},  describes the full set of weighting sectors.

For a graph with in--degree $K=2$, each node has three input links (two regulatory links from other nodes and the constitutive activation/repression), and there is a total of 4 weighting sectors described in the main text (Fig.~\ref{fig:weightSectors}b). Three of these sectors correspond to a combination of weights when one link is stronger than the other two put together, e.g. $w_1>w_2+w_3$. We call these sectors ``dominating''. The three Boolean functions of dominating sectors $\phi^{(a)}$, $a\in\{1,2,3\}$ are given by $\phi^{a}(b_1, b_2, b_3)=b_a$. The Boolean function of the unique non--dominating sector is given by
$$
\phi(b_1,b_2,b_3)=(b_1 \wedge b_2) \vee (b_2 \wedge b_3) \vee (b_3 \wedge b_1).
$$
For a node with weights of incoming links drawn from this sector,  Eq \eqref{app:regulatoryRule} is satisfied whenever any two of the links are satisfied.

For reasons that will be explained shortly, in this work, we consider graphs of topological in--degree $K=3$, which means we have 4 controlling links per node. In this case there are exactly 12 weighting sectors, summarized in Table~\ref{tbl:12sec}: 4 cases with a single ``dominating'' link (one weight is stronger than all others  put together, so the state of this one link defines whether the whole configuration is allowed or not), 4 ``sub-dominating'' (the strongest link and any other must be satisfied, or all three weakest) and 4 ``combinatorial'' (any pair from a given set of three should be satisfied).

\begin{table}[t]
\begin{tabular}{|lc|l|}
\hline
\qquad Sector name &  & Minimal config. of satisfied links\\
\hline
Dominating & $D_a$  & Link $a$ ($a\in\{1,2,3\}$)\\
\hline
Sub-dominating & $SD_a$ & Link $a$ and any other, or all but $a$\\
\hline
Combinatorial & $C_a$ & Any two excluding $a$\\
\hline
\end{tabular}\caption{Weighting sectors for in--degree $K=3$.}\label{tbl:12sec}
\end{table}

\begin{figure}[b]
\includegraphics[width = 0.47 \textwidth]{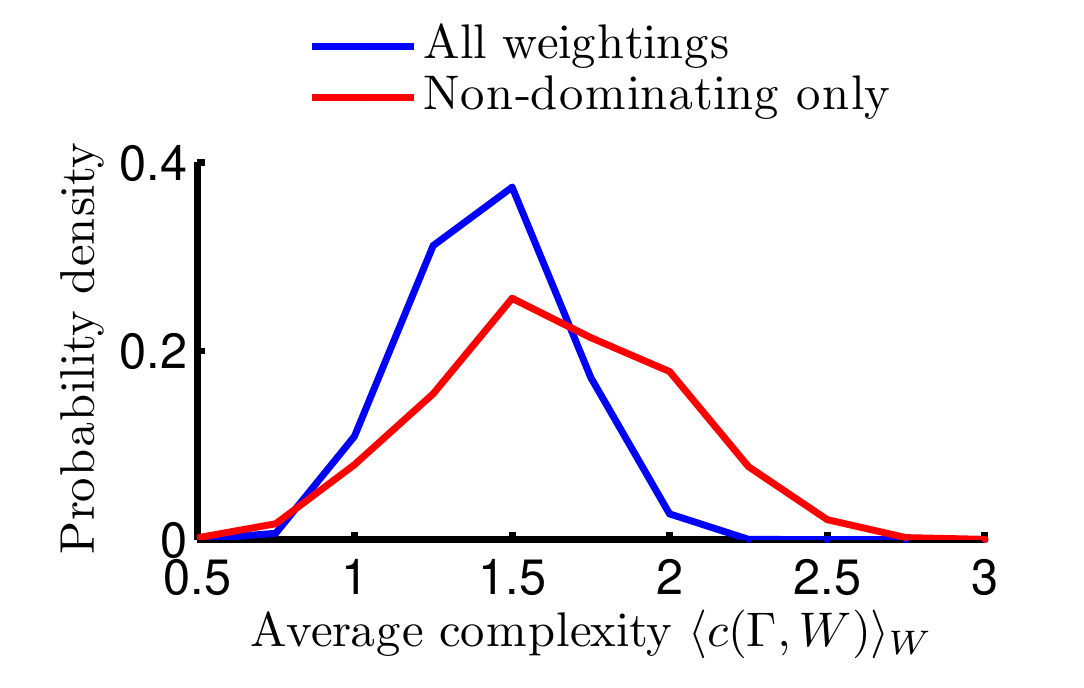}
\caption{Distribution of average complexity of all $N=6$ topologies. Restricting weighting sectors to only non--dominating ones gives a higher average complexity, consistent with expectation that dominating sectors correspond to an effectively simpler topology.}
\label{fig:S1}
\end{figure}

As mentioned in the main text, a node regulated by a dominating link is either redundant with another node or disconnected from the rest of the network (if the dominating weight is the internal threshold). Such nodes cannot increase the complexity of a network, so one expects that forbidding dominating sectors should increase the complexity of a graph. This is indeed correct: Fig.~\ref{fig:S1} shows the distribution of average complexity $\langle c(\Gamma, W)\rangle_W$ over all $N=6$ topologies. Restricting weighting sectors to only non--dominating ones increases the average complexity. This, however, should be seen as the effect of topology rather than parameter choice, because when we picked a topology of in--degree $K=3$, we already recognized the need to have more than 1 regulatory input per node. Therefore, in this work, to distinguish between the effect of weights and topology, we consider non--dominating weighting sectors only. 
With $K=2$ we have only a single non--dominating sector (Fig \ref{fig:weightSectors}b), while with $K=4$ we have  76 (!).  Correspondingly, we studied networks of in--degree $K=3$, as it is the simplest nontrivial case.

It must be noted that in a combinatorial sector $C_a$, $a\in\{1,2,3\}$, the weight of the link $a$ is so weak it has no effect on the state of the target node. For example, a representative set of weights from the sector $C_1$ is $\{1, 5, 5, 5\}$, and removing the first link will not affect the set of solutions of this particular weighting. Despite this fact, the high-complexity graphs are highly enriched in combinatorial sectors, because they in fact perform the most nontrivial local calculation in the information-theoretic sense. To see this, we consider the Boolean input-output functions implemented by a given $K=3$ node and compute the mutual information between any one input $s_{\mathrm{in}}^{(\rm a)}$ and the output node $s_{\mathrm{out}}$, defined as the reduction in entropy of the output brought by the knowledge of the state of a particular input:
\begin{eqnarray}
I(s_{\mathrm{out}}, s_{\mathrm{in}}^{(\rm a)})&=&S[p(s_{\mathrm{out}})]-S[p(s_{\mathrm{out}}\mid s_{\mathrm{in}}^{(\rm a)})]\\
&=&1-S[p(s_{\mathrm{out}}\mid s_{\mathrm{in}}^{(\rm a)})]. 
\end{eqnarray}
Here $S[p(\cdot)]$ denotes the entropy of a probability distribution, and the unconditional entropy of the output is exactly 1 bit. The complexity of the local computation performed by a node can be quantified as the maximum information any one input brings about the output: the lower the information, the more complex the computation. The one performed by a dominating node is trivial: the output is entirely determined by the strongest input (which carries 1 bit). A sub-dominating node fares better: the strongest input carries only
$$
1+\frac 18 \log_2\left(\frac18\right) + \frac 78 \log_2\left(\frac78\right)\approx 0.46 \text{ bits}.
$$
For the combinatorial node, no input carries more than $\approx 0.19$ bits, and thus it performs the most interesting calculation.

Within the constraints of a given $K=3$ topology, a network has the opportunity to regulate each node by up to four inputs. But since the highest-complexity graphs we have found consist exclusively of combinatorial weight sectors, we conclude that to encode the most states, a network may find it should exercise its freedom of parameter choice to only use a particular subset of three inputs at each node.

\section{Volumes of the weighting sectors}
\label{app:IWCvolumes}

As mentioned in the main text, a global rescaling of weights $\vec w \mapsto \alpha \vec w$ at any one node leaves the equation~\eqref{app:regulatoryRule} invariant. Factoring away this global symmetry, the parameter space becomes a sphere or, equivalently, the projective space $\mathbb {RP}_+^{K+1}$. This space being compact, for the microscopic measure $\rho(\vec w)$ in the parameter space we can choose a uniform distribution over this sphere. For in--degree 3 (+ constitutive activation/repression), denoting the four weights as $\{x,y,z,w\}$ and rescaling them so that $x^2+y^2+z^2+w^2=1$, this measure becomes a uniform measure on the 3-dimensional sphere. The relative probabilities of our 12 weighting sectors are then proportional to the volumes they occupy in this space. Direct integration shows that all these weighting sectors  have equal weight. For example, consider the dominating sector $D_1$. Using standard spherical coordinates, we write:
\begin{equation}
V(D_1)=\int_{\Omega} \sin^2(\theta)\sin(\phi)d\theta\,d\phi\,d\psi,
\end{equation}
where $\Omega$ is defined by the conditions $y>0$, $z>0$, $w>0$, and $x>y+z+w$. Rewriting the last condition as
$$\theta<\arccot\left(\cos\phi+\sin\phi\cos\psi+\sin\phi\sin\psi\right)$$
and integrating, we find that this volume is equal to $\pi^2/96$, or $1/12$ of the total volume of the positive sector of the unit 3-sphere.

\section{Computational details}
\label{app:Computation}

Computations were performed with a C++ code, on network topologies with in--degree 3 consisting of repressing interactions only. The choice to only use repressing interactions was motivated by the spin--glass intuition that the diversity of solutions arises from the phenomenon of frustration. This assumption is not overly restrictive, since in our model, an inactive repressor acts as an activator. Note, however, that a network consisting exclusively of activators always trivially possesses the solution $s_{\rm i}\equiv 1$, whereas for repressors, the existence of even one solution is not guaranteed.

Efficient computation was made possible by the following observation. Assume the topology of the graph is fixed. Determining which configurations are solutions of a given weighting is computationally hard. However, our constraints~\eqref{app:regulatoryRule} possess a special structure: there is one local constraint per node, and it involves only the weighting sector associated with this node itself. In other words, the constraints are factorized over the local choice of a weighting sector. This makes the inverse question, ``given a node state configuration $\{s_{\rm i}\}$, for which weights is this a solution?'' extremely simple. Each pattern of satisfied and frustrated incoming links (set by the node states) defines a list of compatible sectors (Table~\ref{tbl:iwcCompatibility}). We can then construct a ``weighting sector compatibility matrix'', a $N\times12$ Boolean matrix identifying, for each node, the list of allowed sectors. For illustration purposes, the dominating sectors are also included in this table; they are not included in the calculations.

The first step of our calculations is always to construct a list of all states that are ever solutions, as well as their compatibility matrices, and takes very little time. These are then used for subsequent steps, to which we now turn.

\begin{table}[t]
\begin{tabular}{|c|c|}
\hline
Link states & Compatible weighting sectors\\
\hline
$\{1,1,1,1\}$ & All \\
$\{1,1,1,0\}$ & All but $D_4$\\
$\{1,1,0,0\}$ & $SD_1, SD_2, C_3, C_4$\\
$\{1,0,0,0\}$ & $D_1$ only\\
$\{0,0,0,0\}$ & None\\
\hline
\end{tabular}\caption{Weighting sectors compatible with a given pattern of satisfied/frusrated links. Dominating sectors included for illustration purposes.}\label{tbl:iwcCompatibility}
\end{table}

\subsection{Targeted search for high-complexity weightings}\label{app:tail}

To determine the distribution of  complexity $c(\Gamma, W)$ over weightings $W$, we perform a recursive tree search by sequentially fixing weight sector choices at every node and calculating, at every step, which subset of the list of potential solutions is compatible with the partially fixed weight sector sequence. The factorization property mentioned above ensures that when choosing a sector at each new node, the list of compatible solutions can only shrink. We can thus perform a targeted search aimed at identifying high-complexity weightings: if we discover early on that the total number of potentially compatible solutions falls below a certain threshold, we can drop the entire subtree of weightings described by the partial specification, and move on. Since most weightings have in fact very few solutions, this allows to exactly enumerate \emph{all} weightings of complexities exceeding a fixed threshold, while still keeping computation time low. If the ``drop-out'' threshold is set to zero, the entire distribution is calculated exactly, and the computation time is still considerably lower than the brute-force enumeration of weightings, taking on the order of a minute on a modern desktop computer for $N=7$. The high-complexity tails can be readily studied up to $N=10$.

To obtain an approximate distribution of solution counts over all weightings for $N=8$-10, when full weighting enumeration is not feasible, we first run a targeted search for high-complexity weightings as described above. We then sample a large number of weightings (e.g., $10^5$) at random and stitch the resulting distribution of low-complexity weightings with the already calculated exact tail of the distribution. This procedure gives excellent agreement for graphs with $N\le7$ where a full distribution can be calculated exactly.

\subsection{Computing the mean complexity of a topology}\label{app:mean}
The average complexity over all weightings $\langle c(\Gamma, W)\rangle_W$ for a given topology $\Gamma$ can be determined without calculating the entire distribution of solution counts by using the following trick. Notice that the average number of solutions is equal to $P(\Gamma)/{N_W}$, where $N_W$ is the number of all possible weightings $W$, and $P=\sum_W c(\Gamma, W)$ is the total number of pairs $(\vec s, W)$, where the state vector $\vec s$ is a solution of the network $(\Gamma, W)$. To calculate $P(\Gamma)$, rather than summing the number of solutions for each weighting $W$, we will sum, for all states $\vec s$, the number $N_W(\vec s)$ of weightings compatible with that state:
\begin{equation}
P(\Gamma)=\sum_W c(\Gamma, W) = \sum_{\vec s}N_W(\vec s).
\end{equation}
To calculate $N_W(\vec s)$ we use the factorization property mentioned above and the Table~\ref{tbl:iwcCompatibility}:
\begin{equation}
N_W(\vec s)=\prod_{\rm i=1}^N N_{\text{sec}}\big[\{f_{\rm j\rightarrow i}(\vec s)\mid{\rm j}\in U({\rm i})\}\big],\label{eq:Ns}
\end{equation}
where $f_{\rm j\rightarrow i}$ denotes, as before, the frustrated/satisfied state of link ${\rm j\rightarrow i}$, and $N_{\text{sec}}$ is the number of weighting sectors compatible with a given pattern of incoming link states (see Table~\ref{tbl:iwcCompatibility}). This trick allows us to rapidly calculate the exact value of typical complexity of graphs up to about $N=25$, and eliminates the need to loop over the weightings themselves, a prohibitively large space for large values of $N$.

For a large $N$, we can also estimate the average complexity $\bar c=\langle c(\Gamma, W)\rangle_{W,\Gamma}$ by making a mean-field approximation in~\eqref{eq:Ns}. For a given node $\rm i$ in the network, each of the incoming links has equal probability of being satisfied and frustrated. Therefore, Table~\ref{tbl:iwcCompatibility} (after excluding the dominating sectors) tells us that the number of allowed weighting sectors at a randomly selected node is a random variable $n$, drawn from a distribution $P$:
\begin{equation}
n=\left[
\begin{aligned}
8&&\text{with probability $5/16$}\\
4&&\text{with probability $6/16$}\\
0&&\text{with probability $5/16$}
\end{aligned}
\right. .
\end{equation}
In the mean--field approximation, we can take this to hold independently for each node in the graph, so
\begin{equation}\label{eq:approx}
\bar c = \frac 1{N_W}\sum_{\vec s} N(\{\vec s\})\approx \frac 1{8^N} 2^N\prod_{\rm i=1}^N n_{\rm i},
\end{equation}
where $n_{\rm i}$ are random variables drawn from $P$. We note that every term in the sum has probability $(11/16)^N$ to be nonzero, and nonzero terms can be rewritten in terms of a new random variable $q$ drawn from $Q=P|_{P>0}$, i.e. from distirbution $P$ conditioned on positivity constraint. We conclude that
$$
\bar c = \left(\frac 2 8 \frac{11}{16}\right)^N \prod_{\rm i=1}^N q_{\rm i}= \left(\frac{11}{64}\right)^N \exp\left(N\langle \ln q\rangle\right)= \alpha^N,
$$
where $\alpha=\frac {11}{64}\exp(\langle\ln q\rangle)\approx 1.01$ is suspiciously close to $1$.

This result can be understood in more general terms. A network in our model is a system of $N$ binary variables, constrained with $N$ binary equations, each forbidding exactly half of the configuration space. One therefore expects, on average, a number of solutions of order $2^N \left(\frac12\right)^N = 1$, irrespectively of $N$. Note that this argument is very general and requires neither the assumption of a constant in--degree nor the fact that genes are modeled as binary variables; it relies only on the fact that the input-output function at each node maps each sets of inputs into exactly one output. The slight difference between the $\alpha$ of the mean-field calculation and $1$ comes from the weak convexity of the logarithm and is most likely within the error of the mean-field approximation. This simple argument shows that for random (uniformly sampled) topologies the average complexity is not expected to increase with $N$. While a large network is in principle capable of storing many patterns (and this can indeed be aided by a biased choice of topology), achieving high complexity requires a careful adjustment of weights.

We would like to contrast this result with the fact that spin glasses can have exponentially many locally stable states \cite{mezard+al_87}. The apparent contradiction comes from the fact that the ground state of a spin glass is not required to satisfy every single node; a ground state only minimizes the frustration in the entire system. In particular, a ground state always exists, in contrast to a ``solution'' of a network in the sense we consider here.

\subsection{Targeted search for high-complexity topologies}\label{app:highCtopo}
\begin{figure}[t]
\includegraphics[width = 0.45 \textwidth]{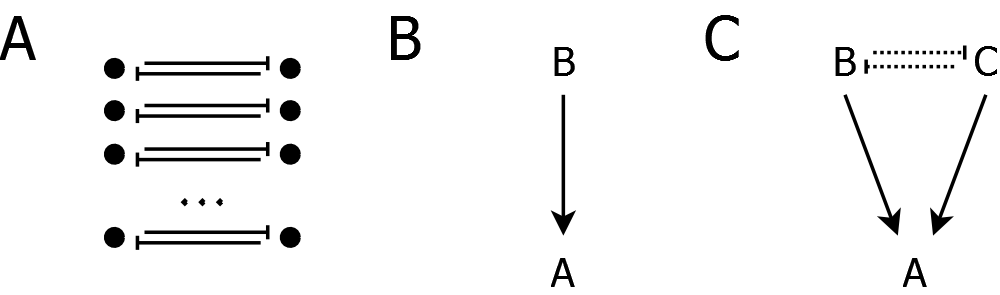}
\caption{\textbf{A:} A simple graph with high capacity. \textbf{B, C:} Two graphs of equal capacity but different regulatory complexity.
\label{fig:diagram}}
\end{figure}

Previous sections discussed techniques to find, for a fixed topology $\Gamma$, the distribution of available complexities $c(\Gamma, W)$ and the optimal weightings that maximize it. Another relevant question to ask is which topologies $\Gamma^*$ achieve the highest complexity, either on average $\langle c(\Gamma^*, W)\rangle_W$ or after optimization of weights $\max_W(c(\Gamma^*, W))$. Finding these topologies becomes particularly important when we realize that disconnected graphs may have a high number of solutions without being truly complex in information-processing sense described in the main text (when the network is seen as performing a mapping between a subset of nodes designated as ``input'' and the rest of the nodes, the ``output''). For example, a disconnected network consisting of $M$ mutually repressing pairs of nodes has $2M$ nodes and can have $2^M$ solutions (Fig.~\ref{fig:diagram}A). In the space of all $N$-node graphs, disconnected topologies are exponentially rare. Therefore, we do not expect disconnected topologies to significantly affect statistical properties, such as average complexity values of all topologies with a fixed $N$. However, to be able to interpret our results, we need to ensure that the highest-complexity networks identified by our measure are not pathological in this manner.

Luckily, the highest-capacity graphs are in fact connected and their capacity exceeds $2^{N/2}$: storing patterns in a network in a distributed way is more efficient than splitting it into many disconnected components. To see this, we need a method to perform a targeted search for high-complexity topologies. Indeed, for $N>6$, there are too many topologies to be sampled exhaustively, and probing tails of heavy-tailed distributions by random sampling is extremely inefficient.

To solve this problem, we used the empirical observation that high-complexity topologies are enriched in mutual-repression motifs (compare Fig.~\ref{fig:N6}, panels A, B and C). The reason for this becomes clear if we recall the mean-field argument we used to calculate the expected number of solutions of a graph. The mean-field approximation assumes, for all links, equal probability of being satisfied or frustrated. In an mutual repression motif, however, the two links are always either both satisfied, or both frustrated. In a topology enriched in mutually repressing pairs, every time a node is satisfied (i.e. has a sufficient number of satisfied incoming links), this increases the probability of their neighbors to be satisfied as well; consequently, such topologies tend to have weightings with larger number of solutions. By sampling only topologies whose $T_{\rm ij}$ matrix (no tilde!) is close to symmetric, we frequently find graphs with complexities far above average. The best networks found in this way have complexity 27 for $N=9$ and 54 for $N=10$, all confidently above $2^{N/2}$. The highest complexity achievable for a given $N$ may be higher still, demonstrating that the highest capacity networks are not trivially so.

We stress that the fact that symmetric $T_{\rm ij}$ matrices lead to higher complexity values is a consequence of our simplifying choice to only consider repressing interactions. Allowing interactions to be both repressing and activating will remove this special structure.

\section{TSP complexity}
\label{app:TSP}

Another approach to deal with pathological cases such as Fig.~\ref{fig:diagram}a is to redefine the complexity measure in a way that does not see such graphs as complex. We will now construct an improved definition of network complexity that quantifies complexity as the diversity of causal relations across the set of solutions. Let us explain what we mean by this, and show that this definition, first, correctly handles pathological cases of disconnected topology, and second, for connected graphs is in excellent agreement with the more simple complexity measure we used in the main text.

We begin with an example. Compare two situations (Fig.~\ref{fig:diagram}b,~c): in the first, gene $A$ is directly regulated by $B$, so they are both ``on'' or both ``off''. In the second, gene $A$ can be activated by either of its two inputs $B$ or $C$, so we can have $A$ ``on'' because $B$ is ``on'', or because $C$ is ``on''. Imagine that $B$ and $C$ are embedded in the network in such a way that they cannot both be active (shown as mutual repression on Fig.~\ref{fig:diagram}c). In this case both networks have two states, but the regulation of $A$ can be described as more ``complex'' in the second case, because the causal relations are more diverse.

\begin{figure*}[tb]
\includegraphics[width = 0.97 \textwidth]{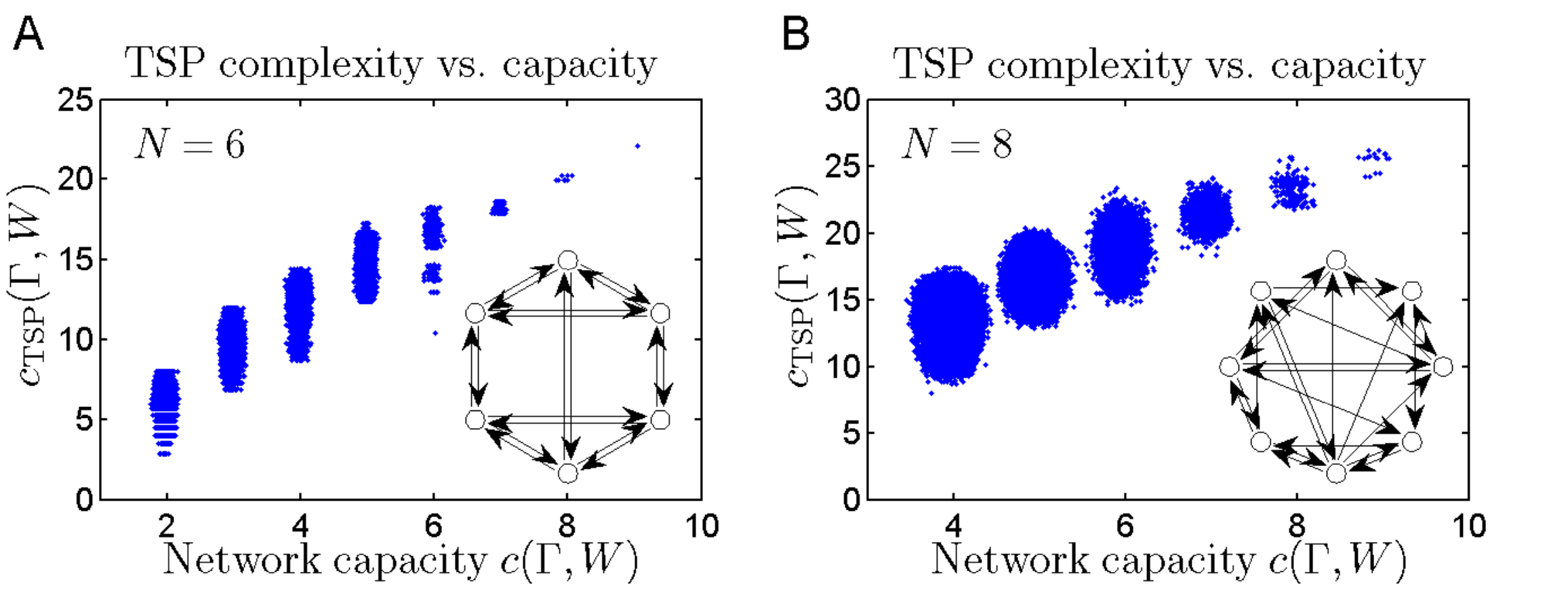}
\caption{For connected graphs, network capacity $c(\Gamma, W)$ is in excellent agreement with the TSP complexity $c_\TSP(\Gamma, W)$. Datapoints correspond to different weightings of the same topology shown in the inset. A small random component was added to the $X$ axis for display purposes. \textbf{A:} the highest-complexity topology of $N=6$ (see Fig.~\ref{fig:N6}C); scatter plot shows all weightings with at least 2 solutions. Weightings with 0 and 1 solutions have zero TSP complexity by definition. \textbf{B:} A random $N=8$ topology with the same maximal capacity (9 solutions). All $2.4 \times 10^6$ weightings with at least 4 solutions are shown. The remaining $1.4\times 10^7$ have 3 solutions or fewer.
\label{fig:S2}}
\end{figure*}

To formalize the intuition gained from this example, we first introduce the notion of an ``active link''. Define a satisfied link to be \emph{active} if its satisfied state is essential for the regulatory rule~\eqref{app:regulatoryRule} to be satisfied. In other words, for a given solution $\{s_{\rm i}\}$, a link $\rm{j_0\rightarrow i_0}$ is ``active'' if and only if substitution $s_{\rm j_0}\rightarrow -s_{\rm j_0}$ upsets equation~\eqref{app:regulatoryRule} at node $\rm i_0$. (Clearly, frustrated links can never be active). Loosely speaking, an active link is ``responsible'' for setting the state of the node $\rm i_0$. Each solution $\vec s$ of a network $(\Gamma, W)$ defines a pattern of active links $\{a^{(\vec s)}_{\rm k}\}$: a binary sequence specifying, for every link $\rm k$, whether it is active ($a=1$) or not ($a=0$). For example, a ``dominating'' link is always the unique active input at the node it regulates, for any solution. For other combinations of link weights, we can have zero, one or more input links active simultaneously, and this pattern will vary from solution to solution.

We now take a moment to define the second ingredient of our definition, the diversity of a set of binary sequences. Given any two sequences, it is easy to define some measure of their difference; we will use the Hamming distance. How can we quantify the diversity of a \emph{set} of sequences? Given a set of elements $G=\{x_1, x_2, \dots, x_k\}$ and a metric of pairwise distances between elements $d$, we would like a reasonable measure of \emph{diversity} $D(G)$ to satisfy three properties:
\begin{enumerate}
\item Invariance under permutation: $D(x_1, x_2, \dots, x_{\rm k})=D(x_{\sigma(1)}, x_{\sigma(2)}, \dots, x_{\sigma(\rm k)})$ for any permutation $\sigma$.
\item Insensitivity to duplication of an element: $D(x, x, y, z, \dots)=D(x, y, z, \dots)$.
\item Additivity: if all $x_{\rm i}$ are pairwise equidistant, $d(x_{\rm i}, x_{\rm j})\equiv d_0$, then $D(x_1, x_2, \dots, x_{\rm k})=kd_0$.
\end{enumerate}
A measure of diversity satisfying all these intuitive properties can be obtained by solving the traveling salesman problem (TSP): $D_\TSP(x_1, x_2, \dots, x_{\rm k})$ is defined as the length of the shortest closed path passing once through each of the ``cities'' $x_{\rm i}$, with distances between cities being defined by the metric $d$.

We now have all the ingredients ready, and define the \emph{TSP complexity} of a graph $\Gamma$ as the TSP diversity of active link patterns across the set of solutions of the network: $c_\TSP(\Gamma,W)\equiv D_\TSP(\{a^{(s)}_{\rm k}\})$. This improved definition naturally accounts for varying pairwise similarity between solutions. In particular, by focussing on the state of links rather than the nodes themselves, we correctly deal with ``simple'' graphs that may have many solutions: note, for instance, that all the solutions of the graph on Fig.~\ref{fig:diagram}a have the exact same pattern of active links, and thus the graph receives zero TSP complexity score despite its large capacity.

In Figure \ref{fig:S2} we show the comparison between $c_\TSP(\Gamma, W)$ and  $c(\Gamma, W)$, and we see that the agreement is quite good. Pathological cases when capacity overestimates true complexity are rare, and are all located at intermediate capacity values, in the bulk of the distribution. Therefore, once again, they neither alter the structure of the high--complexity tail nor affect statistical properties of the distribution significantly. This improved measure of complexity, however, is computationally hard to evaluate, and the computational ``tricks'' discussed above no longer apply. Therefore, for the purposes of this work, we chose to use the simpler definition.


\begin{thebibliography}{99}
\bibitem{boolean}
There is a large literature describing genetic networks in Boolean terms, going back at least to Ref \cite{kauffman_69}.  Following Kauffman, much of this work considers each gene as being controlled by an arbitrary Boolean function of its inputs.  Here, more in the spirit of Ref \cite{hopfield_82}, we take the summation/threshold model to be a coarse--grained version of more microscopic models, as in Ref \cite{bintu}.
\bibitem{noise}
Part of our simplification is to ignore noise, so that Eq (\ref{act2}) is deterministic.  With noise  the symmetry is only approximate, since scaling inputs down to be smaller than the noise will produce qualitatively different behavior.  Most of the literature on network complexity  also considers the noiseless limit, so we feel justified in starting there.  There is a separate literature about design principles related to combating noise and optimizing information transmission in genetic networks  \cite{bialek_12,infomax,snrmax}, but connecting these different views is beyond the scope of this paper.
\bibitem{auto}
In the presence of explicit auto--regulation, the allowed state of a node is no longer a single--valued function of its inputs:  consider, for example, bistability associated with strong auto--activation. The effect of auto--regulation can be implemented in our model as a feedback loop involving an extra node.
\end{thebibliography}
\end{document}